\documentclass[a4paper]{jpconf}
\usepackage{graphicx}
\begin{document}
\title{Longitudinal magnetic field increases critical current
       in superconducting strip}

\author{Ernst Helmut Brandt$^1$ and Grigorii Mikitik$^{1,2}$}

\address{$^1$Max-Planck-Institut f\"ur Metallforschung,
         D-70506 Stuttgart, Germany}
\address{$^2$B.~Verkin Institute for Low Temperature Physics and
         Engineering, Ukrainian Academy of Sciences, Kharkov 61103,
         Ukraine}

\ead{ehb@mf.mpg.de}

\begin{abstract}
  The critical state of the vortex lattice in a thin strip is
considered for the case when first a perpendicular magnetic field
is applied, then a longitudinal field, and then again the
perpendicular field is increased. This longitudinal field can
strongly enhance the critical currents in the strip since the
vortices are inclined and the currents flow in the strip plane.
\end{abstract}

\section{Introduction}

In the usual Bean critical states of type-II superconductors the
critical currents circulate at a right angle to the local magnetic
induction. Such states usually occur when the shape of the
superconductor is sufficiently symmetric and the external magnetic
field ${\bf H}_a$ is applied along a symmetry axis, so that the
direction of the currents is dictated by the symmetry of the
problem. However, in real samples of nonsymmetric shape, or when
the applied magnetic field changes not only in amplitude but also
in its direction, adjacent flux lines may be slightly rotated
relative to each other in the critical state. This rotation
generates a component of the current {\it along} the local
magnetic field, $j_{\parallel}$, and hence the magnetic fields and
currents are not perpendicular to each other. The theory of such
critical states was developed in Refs.~\cite{MB05,crst1} under the
assumption that $j_{\parallel}$ does not exceed a critical value,
such that flux-line cutting does not occur in the sample. As an
example of such unusual critical states, in Refs.~\cite{crst1} we
considered a thin infinitely long strip in the perpendicular
magnetic field $H_z^0$, with this field tilting towards the axis
of the strip. In this paper we shall consider the same example but
we assume now that after the tilt, the perpendicular magnetic
field begins to increase again. We shall show that in this case
critical states with currents by far exceeding the usual critical
value are generated in the sample. Note that this critical state
problem is similar to the situations investigated in
Refs.~\cite{i1,i2}. But in those experiments an initial
perpendicular magnetic field $H_z^0$ was not applied, and this led
to flux-line cutting.

\section{Results}

Consider a thin strip that fills the space $|x|\le w$, $|y|
<\infty $, $|z|\le d/2$ with $w\gg d, \lambda$ where $\lambda$ is
the London penetration depth. Let the magnetic field $H_{az}^0$ be
first applied to the strip, and then the field $H_{ay}$ is
switched on. The critical current density $j_{c\perp}$ is assumed
to be independent of the magnetic induction $B$, and we also imply
that $H_{az}^0$ and $H_{ay}$ considerably exceed $J_c=j_{c\perp}d$
and the lower critical field $B_{c1}$ (thus, $B=\mu_0H$). On applying
$H_{ay}$, we arrive at the critical states that were investigated
in Ref.~\cite{crst1}. In particular, the sheet-current profiles
$J_y(x,H_{az}^0,H_{ay}) \equiv J_0(x)$  take the form
shown in Fig.~1. Below we imply that these $J_0(x)$ are known. The
sheet currents $J_0$ are less than $J_c$ in the obtained critical
states since in these states the current distribution has a
complicated structure: the local currents ${\bf j}(x,z)$ flow at
an angle to the strip axis, and moreover, their direction varies
across the thickness of the sample.

Let at some moment of time $t=t_0$ the field $H_{az}$ begin to
increase again ($H_{az}\ge H_{az}^0$). Then, the increased field
$H_{az}$ penetrates into the sample, and new critical states
develop in the strip. The penetration of $H_{az}$ is similar to
that analyzed in Ref.~\cite{eh2}, but now this process occurs in a
strip in which a {\it nonuniform} distribution of the sheet
current $J_0(x)$ and of the magnetic field $H_z(x)$ generated by
this current plus the applied $H_{az}^0$ existed at $t\le t_0$. We
find the exact solution of the critical state equations that
describe this penetration process. Our solution generalizes that
of Ref.~\cite{eh2}. At $H_{az}\ge H_{az}^0$ the profiles of the
sheet current take the form, Fig.~2:
\begin{eqnarray}\label{1}
 J(x)&=&J_c(H_{az}),~~~~~~~~~~~~~~~~~~~~~~~~~~~~~~~~~~~~
   b\le x \le w, \\
 J(x)&=&{2J_c(H_{az})\over \pi}\arctan \left [{x\over
 w}{\sqrt{w^2-b^2}\over \sqrt{b^2-x^2}} \right ]+J_0(x) \nonumber
     \\
 &-&\!\!{2x\sqrt{b^2-x^2} \over
 \pi}\!\!\int   _{b}^{w}\!\!\!\!\!{J_0(u)du \over
 \sqrt{(u^2-b^2)}\,(u^2-x^2)}, ~~ 0\le x\le b,   \nonumber
\end{eqnarray}
where $J_c(H_{az})=J_c/\cos\theta$,
$\cos\theta=H_{az}/\sqrt{H_{az}^2+H_{ay}^2}$, and $b$, the
position of the penetrating flux front, is obtained from the
implicit equation
\begin{eqnarray}\label{2}
\pi(H_{az}-H_{az}^0) = J_c(H_{az})\ln \left [{w+\sqrt{w^2-b^2}
\over b} \right ]  - \int_{b}^{w}\!\!\!\!\!{J_0(u)du \over
 \sqrt{(u^2-b^2)}}\,.
\end{eqnarray}
At $H_{az}^0=H_{ay}=J_0(x)=0$ equations (\ref{1}) and (\ref{2})
reproduce the solution of Ref.~\cite{eh2}.

The data of Fig.~2 show that the sheet current $J(x)$ sharply
increases with increasing $H_{az}$ and reaches the value
$J_c(H_{az})=J_c/\cos\theta$, which can highly exceed $J_c$.
Knowing the sheet current $J(x)$, one can calculate the magnetic
moment per unit length of the strip, Fig.~3,
     \begin{equation}  \label{3}
  M_z(H_{az}) = -2 \int_{0}^w \! x J(x)\, dx \,,
     \end{equation}
and the profiles $H_z(x)$ for different $H_{az}$ using the Biot -
Savart law, Fig.~4.
Figure 3 also shows $J_c(H_{az})$,  which slightly decreases with
increasing $H_{az}$ due to the decreasing tilt angle $\theta$.

Using the obtained solution and the ideas of Refs.~\cite{ani}, we
also find the distribution of the currents across the thickness of
the strip. In the edge region $b\le x \le w$ one has
$\varphi(x,z)=\pi/2$ where the angle $\varphi(x,z)$ defines the
direction of the local current at a point ($x,z$), ${\bf
j}(x,z)=j_c(\cos\varphi,\sin\varphi,0)$. Here $j_c=j_{c\perp}/[1 -
\sin^2\varphi \sin^2\theta]^{1/2}$  is the magnitude of the
critical current density \cite{crst1}. This magnitude exceeds
$j_{c\perp}$ since the current is not perpendicular to the local
magnetic field. At $\varphi = \pi/2$ this formula gives
$j_c(\theta) = j_{c\perp} / \cos\theta $, and we arrive at the
relation $J_c(H_{az}) = dj_c(\theta)$.
In the inner region $0\le x\le b$ the distribution
$\varphi(x,z)$ is determined by formulas (45), (46) of
Ref.~\cite{crst1}, i.e.,
\begin{eqnarray}\label{4}
 \cot \varphi(x,z)=- {z\cos\theta\over a(x)},
 \end{eqnarray}
where the length $a(x)$ is found from the equation
\begin{equation}\label{5}
\!\!\!\!\!{J(x)\over J_c}\cos\theta\!=\!{2a\over d}\ln\!\! \left
(\!{d\over
  2a}+\sqrt{1+{d^2\over 4a^2}}\,\right )\!\!=\!{2a\over d}\,
  {\rm arsinh} {d\over 2a}\,.
\end{equation}
Note that in this region of the strip the local currents
${\bf j}(x,z)$ flow at different angles to the $y$ axis for
different $z$, and so $J(x) \ne d j_c(\theta)$, see Fig.~5
and \cite{7,8}.

 \begin{figure}  
\includegraphics[width=21.2pc]{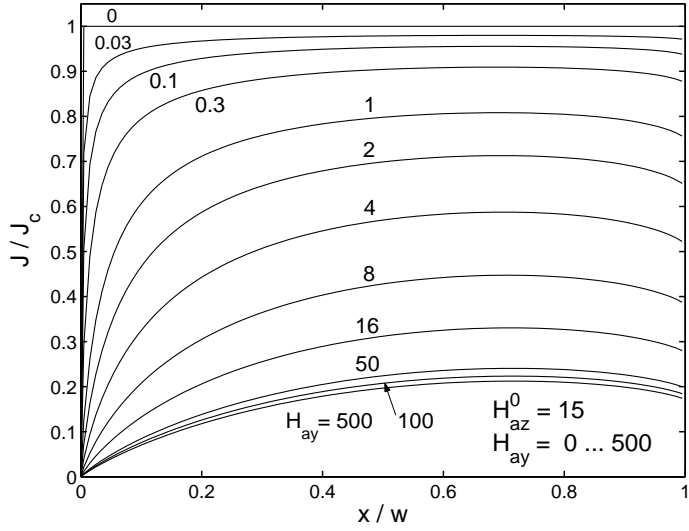}\hspace{2pc}%
\begin{minipage}[b]{15pc} \caption{\label{fig1}
  Sheet current $J_0(x)\equiv |J_y(x)|$ in a strip to which first a
  large perpendicular magnetic field $H_{az}=15$ is applied and
  then an increasing longitudinal field $H_{ay}$, which leads to
  a decrease of the current. The aspect ratio of the strip is
  $2w/d = 20$. The sheet currents and magnetic fields are in
  units of $J_c=j_{c\perp}d$.}
 \end{minipage}
 \end{figure}   

\begin{figure}  
 \includegraphics[width=89mm]{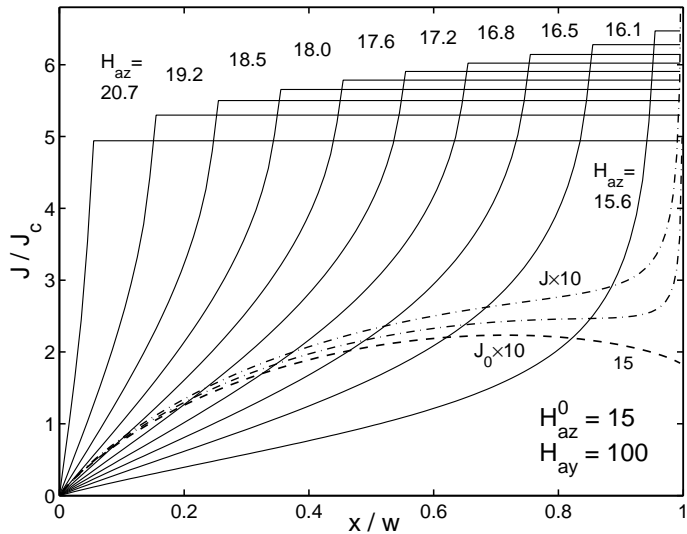}\hspace{2pc}%
\begin{minipage}[b]{15pc} \caption{\label{fig2}
  Sheet current $J(x)\equiv |J_y(x)|$ in a strip as in Fig.~1.
  First a large perpendicular magnetic field $H_{az}^0=15$ is
  applied, then a longitudinal field $H_{ay} =100 $, and
  then $H_{az}$ is increased again. This leads to the
  penetration of a large sheet current
   $J = J_c/\cos\theta \approx 6J_c$. Shown are the penetration
  depths $b/w= 0.95$, 0.85, 0.75, $\dots$, 0.15, 0.05.
  The penetration is complete when $H_{az} \approx 21$
  (units $J_c$). The dashed line shows the much smaller
  initial current $J_0(x)$, the curve $H_{ay}=100$ in Fig.~1.
  The dash-dotted lines show $J(x)$ at fields very close to
  $H_{az}^0$.}
 \end{minipage}
 \end{figure}   

 \begin{figure}  
\includegraphics[width=21.2pc]{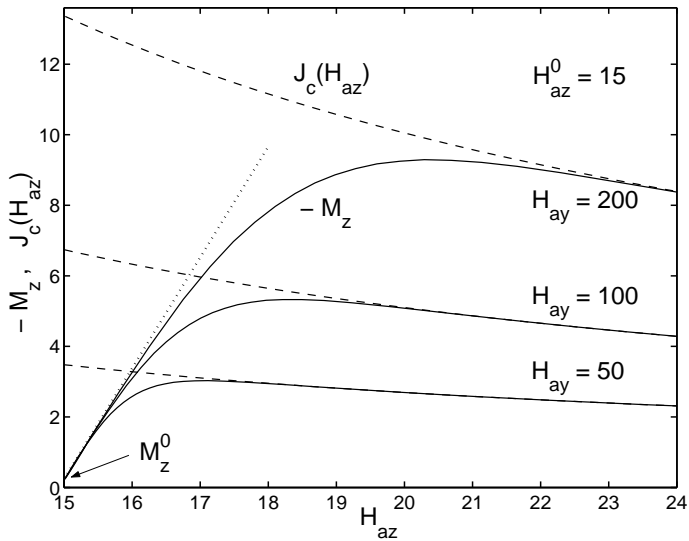}\hspace{2pc}%
\begin{minipage}[b]{15pc} \caption{\label{fig3}
  The magnetic moment $M_z$ (solid lines, units $J_c w^2$) and
  the critical current $J_c(H_{az})$ (dashed lines, units $J_c$)
  for the strip of Fig.~2 with $H_{az}^0=15$, now for three
  values $H_{ay}=50$, 100, and 200, plotted versus the
  increasing $H_{az}$. At $H_{az}=H_{az}^0$ one has a small
  initial $M_z^0$ from $J_c^0$, then $-M_z$ increases with the
  ideal-screening initial slope
  $-\partial M_z / \partial H_{az} =\pi w^2$ (dotted line)
  and reaches a high maximum due to the penetrating large
  $J_c(H_{az})$.
  After full penetration, $J_c(H_{az})$ and $M_z(H_{az})$
  coincide in these units.}
 \end{minipage}
 \end{figure}   

 \begin{figure}  
\includegraphics[width=21.2pc]{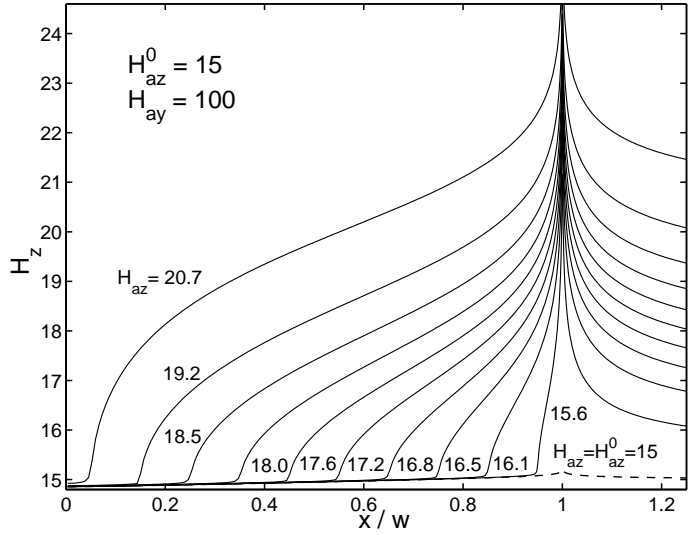}\hspace{2pc}%
\begin{minipage}[b]{15pc} \caption{\label{fig4}
 The magnetic field $H_z$ caused by the sheet current $J(x)$ of
 Fig.~2 and by $H_{az}$ when $H_{az}$ is increased, for same
 penetration depths $b/w=$ 0.95, 0.85, $\dots$, 0.15, 0.05 as
 in Fig.~2. In the central part of the strip $H_z$ initially
 remains frozen at the initial profile
 $H_z^0(x) \approx H_{az}^0$ (dashed line).}
 \end{minipage}
 \end{figure}   

 \begin{figure}  
\includegraphics[width=21.2pc]{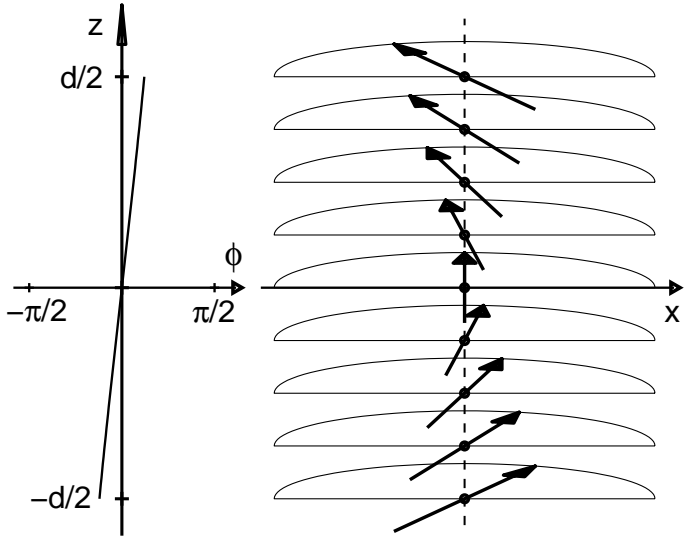}\hspace{2pc}%
\begin{minipage}[b]{15pc} \caption{\label{fig5}
The direction of the local current density ${\bf j}(x,z)$
that flows in the strip plane at various heights $z$
in the central part $|x| \le b$ of the strip.
The angle $\phi = \varphi- \pi/2$ between ${\bf j}(x,z)$
and the strip axis ($y$ axis) is shown at the left
for this example.}
 \end{minipage}
 \end{figure}   

\ack{This work was supported by the German Israeli Research Grant
(GIF) No G-901-232.7/2005.}

\end{document}